\documentclass{kluwer}
\usepackage{graphicx}
\begin{document}
\begin{article}
\begin{opening}
\title{Modeling of $\alpha$\,Cen and Procyon using VLTI observations}            

\author{Pierre \surname{Kervella}\email{Pierre.Kervella@eso.org}}
\institute{European Southern Observatory, Chile}
\author{Fr\'ed\'eric \surname{Th\'evenin}}
\author{Pierre \surname{Morel}}
\author{Gabrielle \surname{Berthomieu}}
\author{Janine \surname{Provost}}
\institute{Observatoire de la C\^ote d'Azur, France}
\author{Pascal \surname{Bord\'e}}
\institute{Observatoire de Paris-Meudon, France}
\author{Damien \surname{S\'egransan}}
\institute{Observatoire de Gen\`eve, Switzerland}

\runningauthor{P. Kervella et al.}
\runningtitle{Modeling of $\alpha$\,Cen and Procyon from VLTI observations}

\begin{ao}
\\ Pierre Kervella\\
European Southern Observatory\\
Alonso de Cordova 3107, Vitacura\\
Santiago, Chile\\
\end{ao} 

\begin{abstract} 
We present a novel approach to model the nearby stars $\alpha$\,Cen A \& B and
Procyon A using asteroseismic and interferometric constraints.
Using the VINCI instrument installed at the VLT Interferometer (VLTI), the angular diameters of
the $\alpha$\,Centauri system were measured with a relative precision of 0.2\% and 0.6\%,
respectively. From these values, we derive linear radii of $R[A] = 1.224 \pm 0.003 R_{\odot}$
and $R[B] = 0.863 \pm 0.005 R_{\odot}$. These radii are in excellent agreement with the
models of Th\'evenin et al.~(\citeyear{thevenin02}), that use asteroseismic frequencies as constraints
(Bouchy \& Carrier~\citeyear{bouchy01}; Bouchy \& Carrier~\citeyear{bouchy02}).
With the same instrument, we also measured the angular diameter of Procyon A. Using the
{\it Hipparcos} parallax, we obtain a linear radius of $2.048 \pm 0.025 R_{\odot}$. We use this result
together with spectroscopic and photometric constraints to model this star with the CESAM
code. We also computed the adiabatic oscillation spectrum of our model of Procyon A, giving
a mean large frequency separation of $\Delta \nu_0 = 54.8\,\mu$Hz, in agreement with the seismic
observations by Mart\`ic et al.~(\citeyear{martic01}). Our model favours a mass around $1.4\,M_{\odot}$ for Procyon A.
\end{abstract}
\keywords{interferometry, numerical modeling, asteroseismology}
\end{opening}

\section{Scientific rationale}
$\alpha$\,Cen~A (G2V) and B (K1V) offer the unique possibility to study the stellar physics at play in conditions
just slightly different from the solar ones. Their masses bracket nicely the Sun's value, while they
are slightly older. In spite of their high interest, proximity and brightness, the two main components
have never been resolved by long baseline stellar interferometry, due to their particularly southern
position in the sky. Their angular diameters were measured recently using the VLTI,
and we have used them to constrain numerical models of these stars.

Procyon is also a binary star, with a white dwarf companion orbiting the main component in 40 years.
Among the brightest stars in the sky, Procyon has been a target for a large number of spectro-photometric
calibration works. However, reproducing its position in the {\sc hr} diagram has been recognized as of great
difficulty (Guenther \& Demarque~\citeyear{guenther93}) using the classical constraints (metallicity, temperature, brightness).
The addition of the asteroseismic and interferometric observational results allows to narrow very
significantly the uncertainties of the evolutionary models.

\section{Interferometric observations}

\begin{figure}
\centerline{\includegraphics[bb=0 0 360 288, width=8cm]{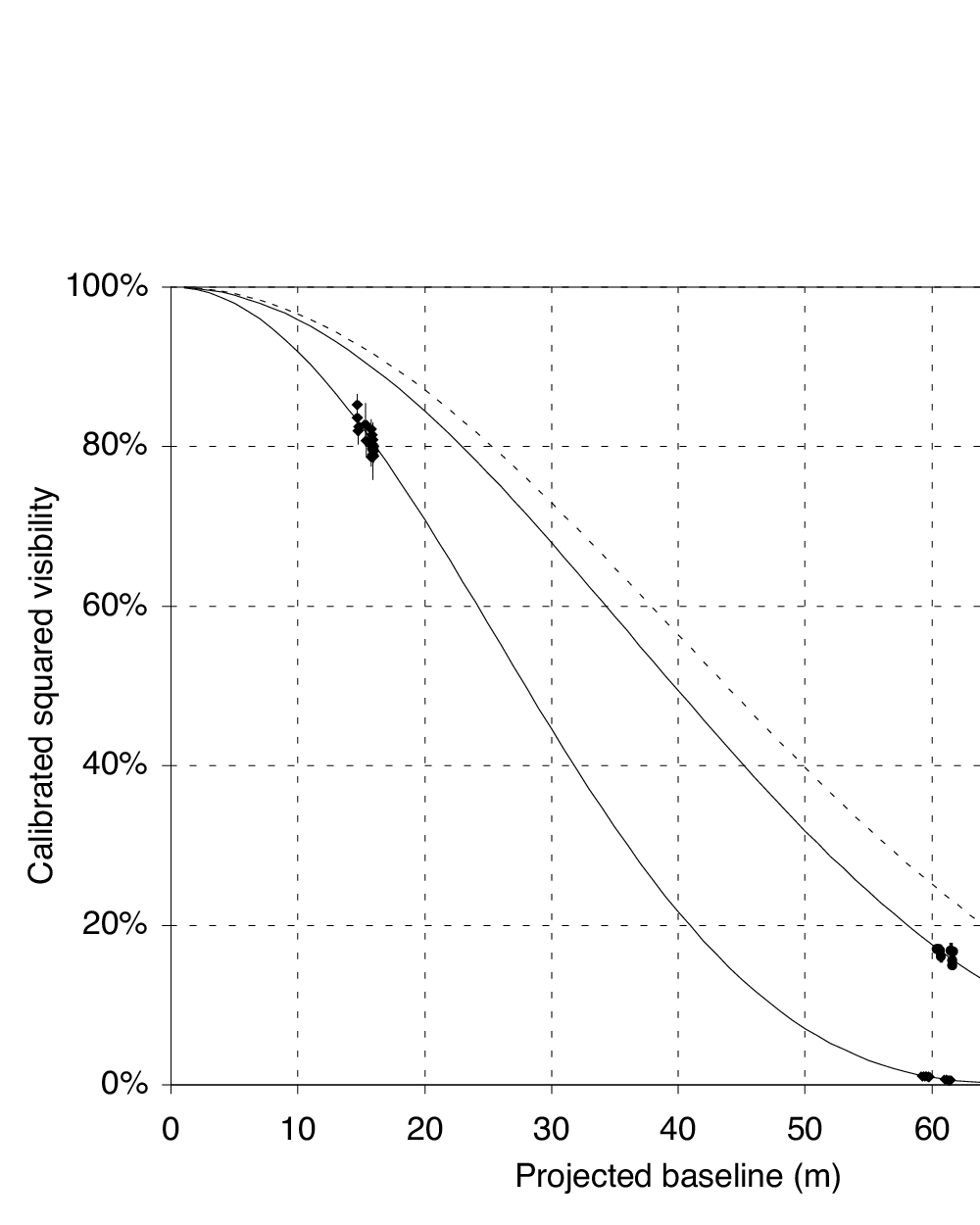}}
\caption{Squared visibilities measured with VINCI on $\alpha$\,Cen A (bottom curve),
$\alpha$\,Cen B (middle curve) and their primary calibrator $\theta$\,Cen (upper curve).}
\label{global_visib}
\end{figure}
\begin{figure}
\centerline{\includegraphics[bb=0 0 360 288, width=8cm]{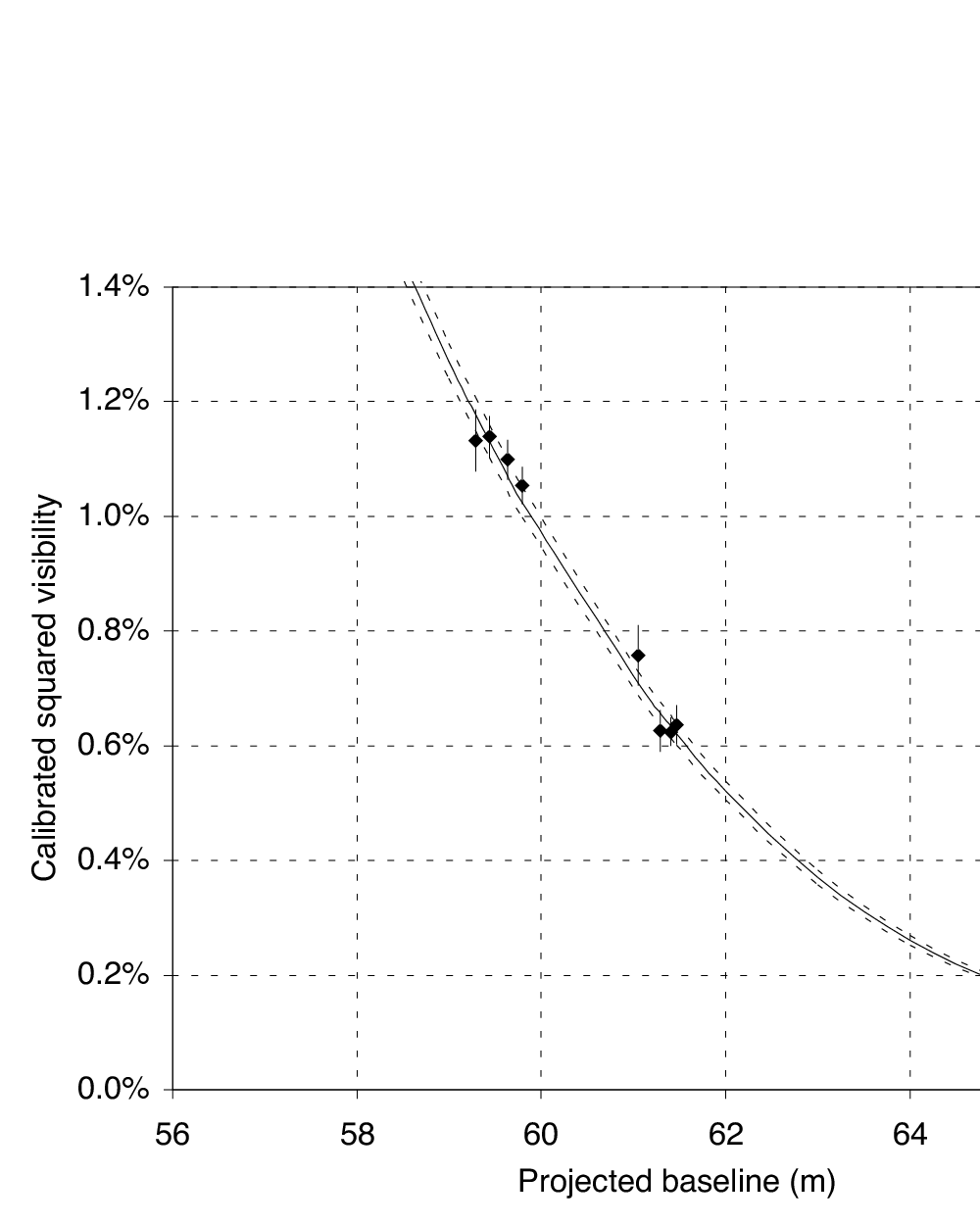}}
\caption{Enlargement of Fig.~\ref{global_visib} showing the low visibility measurements obtained on $\alpha$\,Cen~A.
They constrain the angular diameter model to a precision of $\pm 0.2$\,\% (error domain limited by the dashed lines).}
\label{alfcena_visib}
\end{figure}

For all our interferometric observations, we used the VLT Interferometer with its commissioning
instrument, VINCI (Kervella et al.~\citeyear{kervella03c}), a two telescopes beam combiner operating in the K band
(2.0-2.2 $\mu$m). This instrument measures the squared visibility ($V^2$) of the interferometric fringes. It is
related to the angular diameter of the star through the Zernike-Van Cittert theorem. Fig.~\ref{global_visib} and \ref{alfcena_visib}
illustrate the $V^2$ measurements that we obtained on $\alpha$\,Cen\,A \& B, and the best-fit models that allowed us to derive
their limb darkened angular sizes: $8.511 \pm 0.020$ and $6.001 \pm 0.034$\,mas (Kervella et al.~\citeyear{kervella03a}).
Coupled with the Hipparcos parallax of $747.1 \pm 1.2$\,mas (S\"oderhjelm~\citeyear{soderhjelm99}), this translates into linear
radii of $1.224 \pm 0.003\,R_{\odot}$ and $0.863 \pm 0.005\,R_{\odot}$, respectively.

\begin{figure}[t]
\centerline{\includegraphics[bb=0 0 360 288, width=8cm]{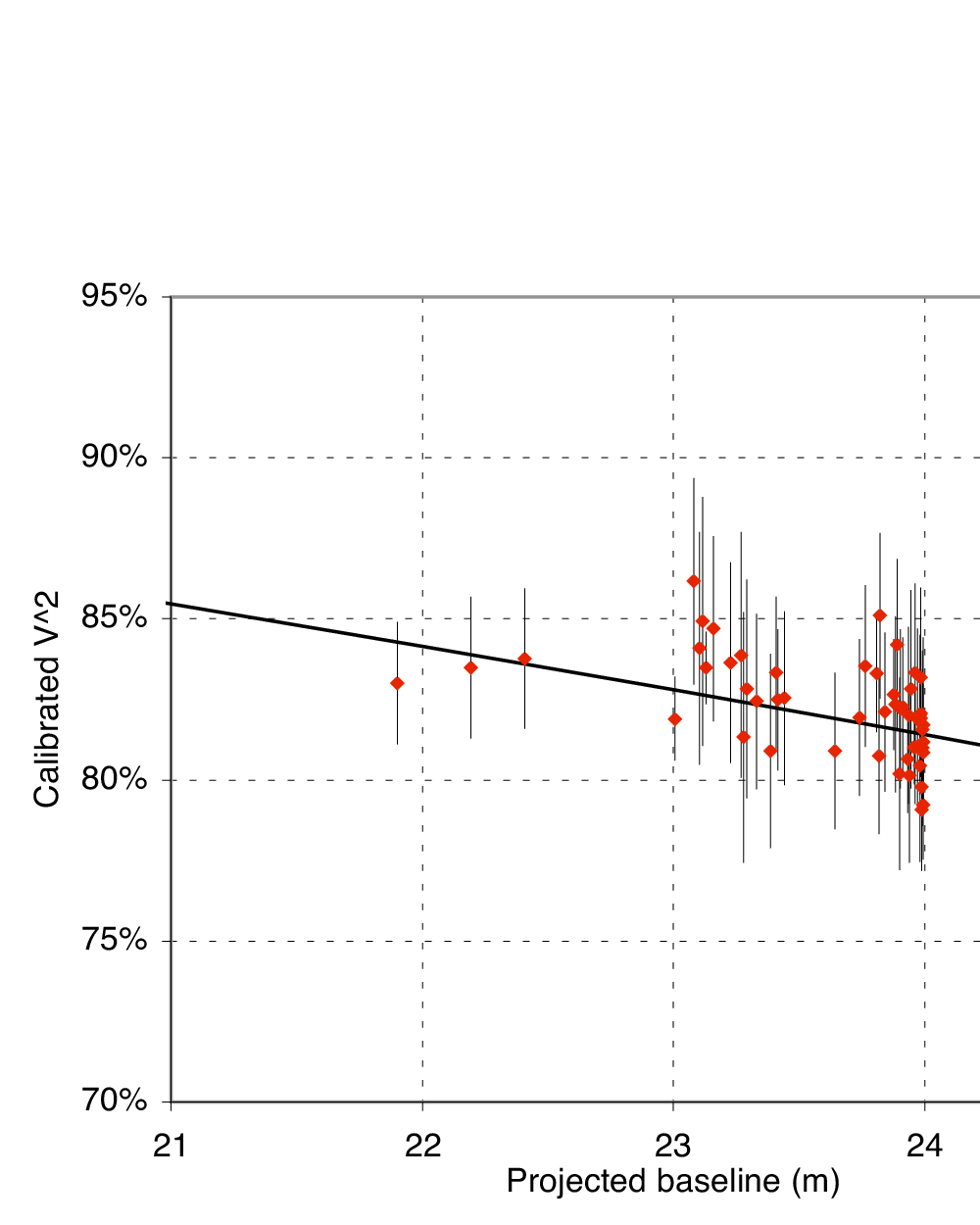}}
\caption{Squared visibility measurements obtained on Procyon with VINCI and best fit model (solid line).}
\label{procyon_visib}
\end{figure}

Using the same method (Fig.~\ref{procyon_visib}), we obtain an angular diameter of
$\theta_{\rm LD} = 5.448 \pm 0.053$\,mas, and a linear photospheric radius of
$2.048 \pm 0.025\,R_{\odot}$ for Procyon~A.

\section{Modeling using the CESAM code}

\subsection{Method}

We have computed a series of models using the CESAM evolutionary code (Morel~\citeyear{morel97}). The evolution
of each star was computed starting from the homogeneous ZAMS corresponding to their pre-estimated
masses. As both $\alpha$\,Cen and Procyon are visual binary stars with well known orbits, it was possible to use
a precise value of their masses for the input of our models. Each model (see e.g. Table~\ref{table_procyon})
corresponds to a different evolutionary track in the {\sc hr} diagram.

\begin{figure}[t]
\centerline{\includegraphics[width=6cm, angle=-90]{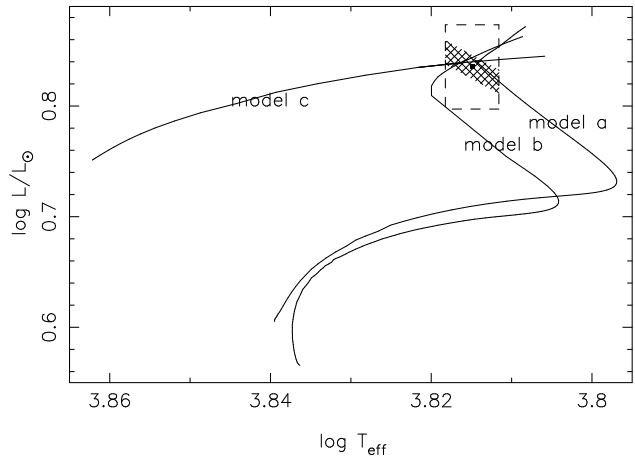}}
\caption{Evolutionary tracks in the {\sc hr} diagram of three models of Procyon\,A (see Table~\ref{table_procyon}
for the corresponding parameters).
The dashed rectangle delimits the uncertainty domain for luminosity and effective temperature,
while the hatched area delimits the uncertainty domain for effective temperature
and the interferometric radius. Model $a$ is the most probable model for this star (see text).}\label{hr_procyon}
\end{figure}

The models were considered acceptable when
their evolutionary track reached the center of the uncertainty domain defined by the photometric, spectroscopic
and interferometric constraints. The constraint imposed by the linear radius $R$ and the effective temperature
$T_{\rm eff}$ is illustrated in Fig.~\ref{hr_procyon} by the hatched parallelogram, while the constraint imposed by $T_{\rm eff}$
and the luminosity $L$ is illustrated by the dashed rectangle. The surface of the shaded area is much smaller than the "classical"
uncertainty domain, and it emphazises the advantage of using the measured $R$ instead of $L$ which depends of
photometric calibrations and bolometric corrections.

\subsection{Procyon}

Procyon was modeled using the parameters listed in Table~\ref{table_procyon}.
Of the models that were tested, our model $a$ succeeds the best in satisfying simultaneously all
the observational constraints.
Model $c$, based on a larger mass of $M=1.50\,M_{\odot}$, 
is clearly rejected by the mean
asteroseismic large frequency spacing $\Delta \nu_0$ (see further for
definition), as its predicted value 56.4\,$\mu$Hz is incompatible
with the value of 54\,$\mu$Hz measured by Mart\`ic et al.~(\citeyear{martic01}).
Model $b$ was computed without microscopic diffusion. It is still marginally
compatible with the observational constraints, and it is older than model a by 400\,Myr.

\begin{table}
\caption[]{Procyon\,A  models (without overshoot) lying within the uncertainty box in the {\sc hr} diagram.
The corresponding evolutions in the {\sc hr} diagram are shown on Fig.~\ref{hr_procyon}.
The subscripts ``$_{\rm i}$'' and ``$_{\rm s}$'' respectively refer
to initial values and surface quantity at present day. ``$_{\rm c}$''
refers to the central value. The model $a$ is the most probable (see text).}\label{table_procyon}
\begin{tabular*}{\maxfloatwidth}{lccc}
\hline
Model                         &     $a$     &     $b$     &     $c$     \\
\hline
\noalign{\smallskip}
$m/M_\odot$                       &   1.42    &   1.42    &   1.50 \\
$Y_{\rm i}$                       & 0.3012 &    0.2580 & 0.345    \\
$Y_{\rm s}$                       & 0.2209 &    0.2580 & 0.202    \\
$\left(\frac ZX\right)_{\rm i}$   & 0.03140 &   0.0218 & 0.0450    \\
$\left(\frac ZX\right)_{\rm s}$   & 0.02157 &   0.0218 & 0.0220   \\
diffusion                         & yes     &   no     & yes \\
$X_{\rm c} $                     & 0.00051 &   0.00000 & 0.2180    \\
age (Myr)                         & 2\,314 &    2\,710 & 1\,300   \\
$T_{\rm eff}$\,(K)                & 6524 &      6547 & 6553    \\
$\log g$                          & 3.960 &     3.967 & 3.994    \\
$\rm [Fe/H]_{\rm i}$              & +0.107 &    -0.051 & +0.264    \\
$\rm [Fe/H]_{\rm s}$              & -0.055 &    -0.051 & -0.043    \\
$\log(L/L_\odot)$                       & 0.8409 &    0.8405 & 0.8390    \\
$ R/R_\odot$                      & 2.0649 &     2.0495 & 2.0420 \\
${\Delta\nu_0}\,(\mu$Hz) & 54.7 &    55.4  & 56.4   \\
\noalign{\smallskip}
\hline
\end{tabular*}
\end{table}

Our model $a$ shows that using the given set of parameters and given physics,
Procyon A is currently finishing to burn its central hydrogen, and is at the 
phase where
the convective core is disappearing.
The error on the measured radius gives a narrow uncertainty of 10 Myr
on the deduced age. Provencal et al. (\citeyear{pro02}) have discussed the cooling 
time
of the WD Procyon\,B and found that the progenitor
ended its lifetime $1.7\pm0.1$\,Gyr ago. 
We derive an age of 2\,314 Myr for Procyon\,A.
Subtracting the cooling age of the WD companion to our determination of the
age of Procyon\,A leads to a lifetime of about 600\,Myr for the progenitor of 
Procyon\,B.
This indicates that the mass of the progenitor is approximately 
2.5\,M$_{\odot}$.
This value yields in turn a mass of $\sim 0.57M_{\odot}$ for the core of the
corresponding Thermal-Pulsating-AGB star (for Z=0.020)
(Bressan et al. \citeyear{bre93}), which is the minimum
possible value for the final mass of the WD (see also Jeffries \citeyear{jeffries}).
This estimate of 0.57\,$M_{\odot}$ agrees very well with the mass of
Procyon\,B that we deduced from Girard et al.\,(\citeyear{girard00}) using
the Hipparcos parallax.
We note that the age obtained with model $c$ is younger than the cooling age of
Procyon~B. This argument suggests a mass lower than 1.5 M$_{\odot}$
for Procyon~A and strengthens the asteroseismology results.

Further progress on the modeling of Procyon will be possible
when the accuracy on the flux of the star is improved to less than 1\,\%.
Waiting for such accuracy, the uniqueness of the solutions resulting from
computed models fitting a narrow box in the {\sc hr} diagram will come from future
detailed asteroseismic studies. For example, other linear combinations of 
frequencies such as the small frequency spacings
(see e.g. Gough \citeyear{go91}) will constrain the age and the mass of the star.

Large uncertainties also come from the adopted chemical 
abundance mixture $Z_s$ which is still rather uncertain.
Only a few chemical element abundances 
are measured today and most of them with a low accuracy. This uncertainty 
on $Z_s$ is the largest source of error on the estimated initial helium content
$Y_i$ and on the age of Procyon.
Thus, we recommend that surface abundances should be derived
from 3D atmosphere studies, in particular for oxygen and other important
donors of electrons.

\subsection{$\alpha$\,Centauri}

As emphasized by Th\'evenin et al. (\citeyear{thevenin02}),
the seismic observations give strong constraints on masses and
on the age of the system when combined with spectro-photometric measurements.
To achieve this, one derives from the set of oscillation frequencies, 
one ``large'' and two ``small'' frequency spacings.
The large frequency spacing is a difference
between frequencies of modes with consecutive radial order $n$:
$\Delta\nu_\ell(n) \equiv \nu_{n, \ell} - \nu_{n-1,\ell}.$ 
In the high frequency range, i.e. large radial orders, 
$\Delta\nu_\ell$ is almost constant with  a mean value $\Delta \nu_{0}$, strongly
related to the mean density of the star, i.e. to the mass and the radius.
The small separations are very sensitive to the physical conditions in the core of the star
and consequently to its age.
These frequencies measured  for the star $\alpha$\,Cen~A have led to decrease
the masses of the stellar system, leading to the following values:
$M_{\rm A}=1.100\pm0.006\,M_\odot$
and $M_{\rm B}=0.907\pm0.006\,M_\odot$ (Th\'evenin et al. \citeyear{thevenin02})
close to those adopted by Guenther \& Demarque~(\citeyear{guenther00}).
The mass of the B component departs significantly by 3\% from the value published by
Pourbaix et al.~(\citeyear{pourbaix02}).

Using radial velocity measurements,
Pourbaix et al. (\citeyear{pourbaix02}) have derived the masses of each components
($M_{\rm A}=1.105 \pm 0.007\,M_\odot$, $M_{\rm B}=0.934 \pm 0.006\,M_\odot$).
We note that Thoul et al. (\citeyear{thoul03}) have recently proposed a model of the
binary system using these masses and spectro-photometric constraints different from that of
Th\'evenin et al.~(\citeyear{thevenin02}). They were able to reproduce the seismic frequencies
of $\alpha$\,Cen\,A, but the model they propose does not take into account the
helium and heavy elements diffusion.

\begin{figure}
\centerline{\includegraphics[width=6cm, angle=-90]{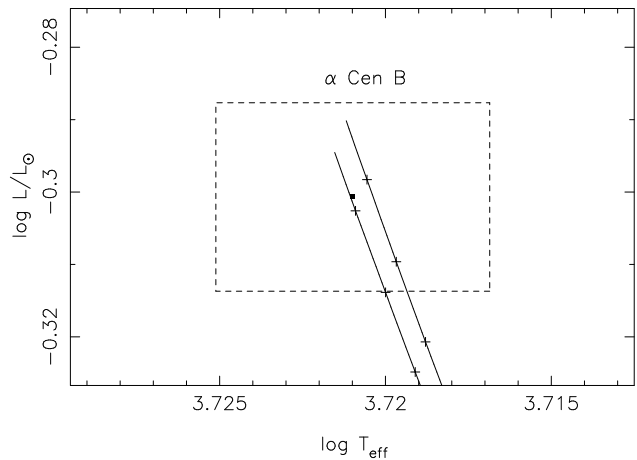}}
\caption[]{Evolutionary tracks of $\alpha$\,Cen~B in the {\sc hr} diagram. The line on the right corresponds to a mixing
length of $\lambda=0.96$ and $M=0.909\,M_{\odot}$, the line on the left corresponds to the values published by
Th\'evenin et al.~(\citeyear{thevenin02}): $\lambda=0.98$, $ M=0.907\,M_{\odot}$.}
\label{hr_alfcen}
\end{figure}

The masses derived by Th\'evenin et al.~(\citeyear{thevenin02}), $M_A=1.100 \pm 0.006$
and $M_B=0.907 \pm 0.006\,M_{\odot}$ allowed us to compute evolutionary models
that match simultaneously all constraints, including the linear diameter
and asteroseismic large frequency spacing. 
These modifications do not change the calibration of $\alpha$\,Cen\,A.
We took care in this process to keep the star B in its error box on the
{\sc hr} diagram (Fig.~\ref{hr_alfcen}). It results from this new mass a diameter
that is closer to the interferometric one: 
0.863 $D_{\odot}$ or $5.999 \pm 0.050$ mas (parallax from S\"oderhjelm~\citeyear{soderhjelm99}).
The effective temperature is found to be 5262 K, identical to the adopted
spectroscopic value $\rm T_{\mathrm{eff}} = 5260$ K.
Our results confirm that the mass of the B component is probably close to 0.907 M$_{\odot}$,
as reported by Th\'evenin et al.~(\citeyear{thevenin02}).

As shown on Fig.~\ref{hr_alfcen}, it is possible to refine the agreement by changing slightly the hypothesis
of the model, in particular for B. The future availability of the large frequency spacing of $\alpha$\,Cen~B will complete the
calibration of the system. While B is still on the Main Sequence with an hydrogen fraction at center of $X_{\rm center}=0.43$,
$\alpha$\,Cen~A is currently near the end ($X_{\rm center}=0.18$), owing to its larger mass.
This also explains its larger diameter compared to the lower Main Sequence M-R relation (Kervella et al.~\citeyear{kervella03d}).

%
%
%
\theendnotes

\end{article}
\end{document}